\documentclass[12pt]{article}

\usepackage{lineno,hyperref}
\usepackage{mathrsfs}
\usepackage[T1]{fontenc}
\usepackage[utf8]{inputenc} 
\usepackage[english]{babel}
\usepackage{amsmath,amsfonts,amssymb}
\usepackage{cite}
\usepackage{graphicx}
\usepackage{braket}
\usepackage{float}

\title{KvN mechanics approach to the time-dependent frequency harmonic oscillator}

\author{Irán Ramos-Prieto, Alejandro R. Urzúa-Pineda\\
	Francisco Soto-Eguibar, H\'ector M. Moya-Cessa\\
	{\small Instituto Nacional de Astrof\'{\i}sica, \'Optica y Electr\'onica, INAOE} \\
	\small {Calle Luis Enrique Erro 1, Santa Mar\'{\i}a Tonantzintla, San Andrés Cholula, Puebla, 72840 Mexico}\\
	$^*$\small {Corresponding author: feguibar@inaoep.mx}}	

\begin{document}
\maketitle

\begin{abstract}
Using the Ermakov-Lewis invariants appearing in KvN mechanics, the time-dependent frequency harmonic oscillator is studied. The analysis builds upon the operational dynamical model, from which it is possible to infer quantum or classical dynamics; thus, the mathematical structure governing the evolution will be the same in both cases. The Liouville operator associated with the time-dependent frequency harmonic oscillator can be transformed using an Ermakov-Lewis invariant, which is also time dependent and commutes with itself at any time. Finally, because the solution of the Ermakov equation is involved in the evolution of the classical state vector, we explore some analytical and numerical solutions.
\end{abstract}

\section{Introduction}
In 1931, Koopman and von Neumann (KvN) formulated a way to get the realm of classical mechanics in terms of operators and state vectors over a Hilbert space \cite{Koopman,vonNeumann}. The development of KvN mechanics provides an operational language that is intimately linked to quantum theory, and in this approach the dynamics in phase space is determined by the probability distribution function $\Gamma(x,p;t)$, which is the  square module of the classical wave functions $\psi(x,p;t)$ (or KvN wave functions) \cite{D_Mauro2002,Gozzi_2002,D_Mauro2003,D_Mauro_Thesis,D_Mauro2004}.

The operational language in KvN mechanics, that underlines a Hilbert space of complex and square integrable functions, has been used to obtain a new approach and reformulation of classical and quantum theories \cite{D_Mauro_New_2003,Gozzi_New_2005,Gozzi_New_2006,Gozzi_New_2010,Gozzi_New_2011,Rivers_New_2012,Gozzi_New_2013,Rajagopal_New_2016}.  Also in this sense, a hybrid mechanics has been proposed, where one can infer classical and quantum dynamics of statistical ensembles of a single particle in one dimension, which is denoted as operational dynamic modelling \cite{Bondar_ODM}. In this theoretical framework, the phase space plays a fundamental role in the description and interpretation of quantum, classical and hybrid phenomena, encompassing open and closed systems \cite{Bondar_Wigner,Bondar_2015,Bondar_2016_Dirac,Bondar_2016_Lindblad,PhysRevLett.119.170402}.

It is well known that the harmonic oscillator is one of the more (if not the most) studied physical model in physics and the time-dependent parameter (mass and frequency) cases arise as an useful extension for explaining dynamical phenomena. The classical and quantum harmonic oscillator has been studied with time dependent mass and frequency \cite{Solimeno,Ermakov_1,Ermakov_2,Caldirola,Kanai,Manko,Vergel}, and in particular a method of exact invariants has been used to give a solution \cite{Moya_Guasti_1,Moya_Guasti_2,Moya_Guasti_3,Iran_Mass_2017}. There exist many physical systems where it is possible to find the time-dependent frequency harmonic oscillator, such as ions on Paul traps oscillating in one dimension with time dependent frequency \cite{Paul,Brown,Cirac,Moya-Cessa2012}, radiation fields that propagate in time dependent dielectric media \cite{Agarwal}, etc.

Lewis and Reisenfeld were the first to use invariant methods to solve Schrödinger equation for a time-dependent frequency harmonic oscillator \cite{Ermakov_1,Ermakov_2}, and more recently this approach has been used, in conjunction with squeeze transformations, in the Ermakov equation to get a closed solution \cite{Moya_Guasti_3,Iran_Mass_2017}. The use of KvN mechanics and operational dynamic modelling applied to statistical copies that behave like a time-dependent frequency harmonic oscillator, provide an operational treatment where the invariants method can be used. In this article, we use the operative language of operational dynamic modelling in order to use the invariants method to solve a Schrödinger-like equation, where the Hamiltonian is replaced by the Liouvillian. 

\section{KvN mechanics}
In this Section, we summarize some basic concepts of the KvN mechanics. Concepts that are intimately related to quantum theory, such as Hilbert space, state vectors and operators; for state vectors and operators we use Dirac notation, thus they are denoted by $\mathscr{H}$ and $\braket{bra|ket}$, respectively.

In every theory dealing with the statistical properties of a system, finding the probability density function $\Gamma(x,p;t)$ is the main goal; thus, our first requirement is the acquisition of the density operator $\hat{\mathbf{P}}$  at time $t$. In the abstract approach of KvN mechanics, we have the following postulates:
\begin{itemize}
	\item The system is defined by a state vector $\ket{\psi}$ that belongs to a Hilbert space $\mathscr{H}$; we ask the state vector to be normalized, so
	\begin{equation}
	\braket{\psi|\psi}=1.
	\end{equation}
	\item The expected value at time $t$ of an observable $\hat{O}$ is $\braket{\psi(t)|\hat{O}|\psi(t)}$.
	\item The probability that at time $t$ the measurement of an observable $\hat{O}$ yields $O$ is $|\braket{\hat{O}|\psi(t)}|^2$.
\end{itemize}
As consequence of Stone's theorem \cite{Stone}, the equation that describes the evolution of the state vector  $\ket{\psi(t)}$ is
\begin{equation}\label{Stone}
i\frac{\partial}{\partial t}\ket{\psi(t)}=\hat{L}\ket{\psi(t)},
\end{equation}
where the operator $\hat{L}$ can be recognized as the Liouvillian or Hamiltonian according to the commutation relation between the position and momentum operators \cite{Bondar_ODM}.\\ 
The probability distribution function $\Gamma(x,p;t)$ is given by $\Gamma(x,p;t)=|\braket{x,p|\psi(t)}|^2$, and the eigenstates of the position and momentum operators form an orthonormal and complete set, according to
\begin{align}
\begin{split}
\hat{x}\ket{x,p}=x\ket{x,p}&,\quad\hat{p}\ket{x,p}=p\ket{x,p},\\
\braket{x',p'|x,p}=\delta(x'-x)\delta(p'-p)&,\quad\int dx dp\ket{x,p}\bra{x,p}=1,
\end{split}
\end{align}
since classically position and momentum operators commute $[\hat{x},\hat{p}]=0$. 

Once we accept and recognize these postulates, we can find the explicit form of the operator $\hat{L}$. For this task, in the framework of the Ehrenfest theorem \cite{Ehrenfest}, we assume multiple copies of a single particle subjected to a potential $U(\hat{x};t)$, next we apply the postulates mentioned above to this statistical set and we get the following system of equations
\begin{align}\label{Ehrenfest}
\begin{split}
\frac{\partial}{\partial t}\braket{\psi(t)|\hat{x}|\psi(t)}&=\braket{\psi(t)|\hat{p}|\psi(t)},\\
\frac{\partial}{\partial t}\braket{\psi(t)|\hat{p}|\psi(t)}&=\braket{\psi(t)|-U'(\hat{x};t)|\psi(t)},
\end{split}
\end{align}
where we have considered an isolated system and $m=1$. Therefore, applying the equation \eqref{Stone} to the previous equations, we get the commutation relations
\begin{align}\label{commutaion_L}
\begin{split}
i[\hat{L},\hat{x}]&=\hat{p},\\
i[\hat{L},\hat{p}]&=-U'(\hat{x};t).
\end{split}
\end{align}
But in classical mechanics the position and momentum operators can be measured with arbitrary precision; i.e., $[\hat{x},\hat{p}]=0$, and the operator $\hat{L}$ can not be a function only of $\hat{x}$ and $\hat{p}$. This implies that it is necessary to stick to it a pair of operators that satisfy the following commutation rules
\begin{align}
\begin{split}
[\hat{x},\hat{\lambda}_x]&=[\hat{p},\hat{\lambda}_p]=i,
\end{split}
\end{align}
which leads to the explicit form of the operator $\hat{L}$,
\begin{equation}\label{Liouvilliano}
\hat{L}=\hat{p}\hat{\lambda}_x-U'(\hat{x};t)\hat{\lambda}_p.
\end{equation}

As we postulated above, the probability distribution function is  $\Gamma(x,p;t)=|\braket{x,p|\psi(t)}|^2$ and to find the dynamics associated with it, it is necessary to use equations \eqref{Stone} and \eqref{Liouvilliano}, and project onto an orthonormal and complete set of vectors $\ket{x,p}$, in such a way that we obtain
\begin{equation}\label{Liouville_equation}
\frac{\partial}{\partial t}\Gamma(x,p;t)=\left[ -p\frac{\partial}{\partial x}+U'(x;t)\frac{\partial}{\partial p}\right] \Gamma(x,p;t),
\end{equation}
where 
\begin{align}
\begin{split}
\hat{\lambda}_x\ket{x,p}=-i\frac{\partial}{\partial x}\ket{x,p},& \quad\hat{\lambda}_p\ket{x,p}=-i\frac{\partial}{\partial p}\ket{x,p}.
\end{split}
\end{align}
The relation \eqref{Liouville_equation} is the Liouville equation of classical statistical mechanics, which provides the evolution of $\Gamma(x,p;t)$ in phase space. It is worth to remark that the addition of an arbitrary function of  position and momentum operators, $f(\hat{x},\hat{p})$, to the right hand side of equation \eqref{Liouvilliano} does not affect the obtained equation of motion, equation \eqref{Liouville_equation}; in other words, there is an invariance in the KvN theory similar to a gauge invariance. This freedom in the theory has been used recently to model a quantum-classical hybrid \cite{FrancoisGay-Balmaz2018}.

It is also important to mention that this conceptual development, altogether with the previous postulates, is known as operational dynamical modelling. In this way, we can infer classical and quantum dynamics; moreover, it is possible to find their unification  \cite{Bondar_ODM,Bondar_Wigner}.

\section{Invariant of the time-dependent frequency harmonic oscillator in KvN mechanics}
Let us consider a set of identical copies of a single particle subject to a time dependent quadratic potential, which leads to a time- dependent frequency harmonic oscillator in KvN mechanics with potential $U(\hat{x};t)=k(t)\hat{x}^2/2$. Under these considerations equation \eqref{Stone} can be written as
\begin{align}\label{psi_t}
i\frac{\partial}{\partial t}\psi(x,p;t)=\left[ \hat{p}\hat{\lambda}_x-k(t)\hat{x}\hat{\lambda}_p\right] \psi(x,p;t),
\end{align}
where it is evident that the Liouvillian $\hat{L}$ depends on time. However, it is possible to show that the above equation has an invariant of the form (see Appendix A)
\begin{align}\label{inviarnte}
\hat{I}=\frac{1}{2}\bigg[\frac{\hat{x}^2}{\rho^2}+(\dot{\rho}\hat{x}-\rho\hat{p})^2+\frac{\hat{\lambda}_p^2}{\rho^2}+(\dot{\rho}\hat{\lambda}_p+\rho\hat{\lambda}_x)^2\bigg],
\end{align}
where $\rho$ obeys the Ermakov equation
\begin{equation}\label{rho}
\ddot{\rho}+k(t)\rho=\frac{1}{\rho^3}.
\end{equation}
It is important to mention that there is a symbiotic relationship with the solution of the time-dependent frequency harmonic oscillator differential equation
\begin{equation}\label{u}
\ddot{u}+k(t)u=0,
\end{equation}
because we can relate $u(t)$ and $\rho(t)$ as follows
\begin{equation}\label{u-rho}
u=\rho\cos\omega_\rho(t)\quad\mbox{and}\quad\rho=u\sqrt{1+\omega_u^2(t)},
\end{equation}
where
\begin{equation}\label{omega_x}
\omega_y(t)=\int\frac{dt}{y^2(t)}, 
\end{equation}
with $y=\rho,u$. \\
It is important to remark that the operators $\hat{\lambda}_x$ and $\hat{\lambda}_p$ are not associated with any physical observable, since all the observables of a classical system are represented by commuting operators, and this makes it impossible to measure (or observe) the invariant given by the equation \eqref{inviarnte}. It is also worth to notice that the obtained invariant, equation \eqref{inviarnte}, depends on the auxiliary unobserved operators $\hat{\lambda}_x$ and $\hat{\lambda}_p$, and that implies that the invariant is sensitive to the phase of the KvN wave function. In our case, the choice we have made is equation \eqref{Liouvilliano}; i.e., the function $f(\hat{x},\hat{p})$ that can be arbitrarily added to the right hand side of equation \eqref{Liouvilliano} has been taken as zero.

We can move to a scenario determined by the following unitary transformations 
\begin{align}\label{T_1T_2}
\begin{split}
\hat{T}_1(t)&=\exp\left[  {i\frac{\dot{\rho}(t)}{\rho(t)}\hat{x}\hat{\lambda}_p}\right]  ,\\ \hat{T}_2(t)&=\exp\left[i\frac{\ln\rho(t)}{2}(\hat{x}\hat{\lambda}_x+\hat{\lambda}_x\hat{x})\right] 
\exp\left[ -i\frac{\ln\rho(t)}{2}(\hat{p}\hat{\lambda}_p+\hat{\lambda}_p\hat{p})\right],
\end{split}
\end{align}
where equation \eqref{psi_t} can be written as 
\begin{align}\label{eq_psi_t_T}
i\frac{\partial}{\partial t}\Psi(x,p;t)=\frac{1}{\rho^2(t)}\big(\hat{p}\hat{\lambda}_x-\hat{x}\hat{\lambda}_p\big)\Psi(x,p;t),
\end{align}
being $\Psi(x,p;t)=\hat{T}_2(t)\hat{T}_1(t)\psi(x,p;t)$. Hence, the solution in the original framework is
\begin{equation}\label{psi_t_T}
\psi(x,p;t)=\hat{T}_1^\dagger(t)\hat{T}_2^\dagger(t)
\exp\left[-i\omega_\rho(0,t)(\hat{p}\hat{\lambda}_x-\hat{x}\hat{\lambda}_p) \right] 
\hat{T}_2(0)\hat{T}_1(0)\psi(x,p;0)
\end{equation}
with
\begin{equation}
\omega_\rho(0,t)=\int_0^t\frac{dt'}{\rho^2(t')}.
\end{equation}
Therefore, the classical wave function will be given by (see Appendix B)
\begin{align}\label{psi_t20}
\psi(x,p;t)=\psi\bigg[x\eta_1(t)-p\eta_2(t),x\eta_3(t)+p\eta_4(t);0\bigg]
\end{align}
where
\begin{align}\label{relations_rho}
\begin{split}
\eta_1(t)&=\frac{\rho_0}{\rho(t)}\cos\omega_\rho(0,t)+\dot{\rho}(t)\rho_0\sin\omega_\rho(0,t)
,\\
\eta_2(t)&=\rho(t)\rho_0\sin\omega_\rho(0,t),\\
\eta_3(t)&=\bigg[\frac{1}{\rho(t)\rho_0}+\dot{\rho}(t)\dot{\rho}_0\bigg]\sin\omega_\rho(0,t)+\bigg[\frac{\dot{\rho}_0}{\rho(t)}-\frac{\dot{\rho}(t)}{\rho_0}\bigg]\cos\omega_\rho(0,t),\\
\eta_4(t)&=\frac{\rho(t)}{\rho_0}\cos\omega_\rho(0,t)-\rho(t)\dot{\rho}_0\sin\omega_\rho(0,t),
\end{split}
\end{align}
being $\rho_0$ and $\dot{\rho}_0$ the initial conditions of the Ermakov equation.

Equation \eqref{eq_psi_t_T} contains a Liouville operator, that although depends on time, commutes with itself at any time $t$; i.e., the Liouville operator is diagonalizable. On the other hand, as along as there is a solution of the  Ermakov equation \eqref{rho}, expression \eqref{psi_t_T} can be used to calculate any observable of the system.

\subsection{Hyperbolically and quadratically growing frequency}\label{s3.1}
The solution of the Ermakov equation and the time-dependent frequency harmonic oscillator are intimately connected, as equation \eqref{u-rho} shows. As an explicit example, we consider the time dependent frequency $k(t)=2\beta^2/\cosh^2(\beta t)$, and as immediate consequence of equation \eqref{u-rho}, we get the solutions
\begin{equation}
u=\tanh(\beta t),\quad\quad\rho=\tanh(\beta t)\sqrt{1+\frac{[\beta t -\coth(\beta t)]^2}{\beta^2}}.
\end{equation}
On the other hand, it is very easy to show the function $\omega_\rho(0,t)$ contained in the state vector \eqref{psi_t_T} complies with
\begin{align}\label{omega_0_t}
\begin{split}
\omega_\rho(0,t)=\int_0^t\frac{dt'}{\rho^2(t')}=\arctan\bigg[\int\frac{dt'}{u^2(t')}\bigg]\bigg|_0^t.
\end{split}
\end{align}
Another example, where it is possible to find analytic relationships, is the one with   frequency $k(t)=1/(\gamma+2t)^2$. Following the recipe shown above, we can find that
\begin{equation}
u(t)=\sqrt{\gamma+2t},\quad\rho(t)=\sqrt{(\gamma+2t)\bigg[1+\frac{1}{4}\ln^2(\gamma+2t)\bigg]},
\end{equation}
and following the relationship \eqref{omega_0_t}, we can find also $\omega_\rho(0,t)$. 

\begin{figure}[H]
	\centering
	\includegraphics[width=16cm]{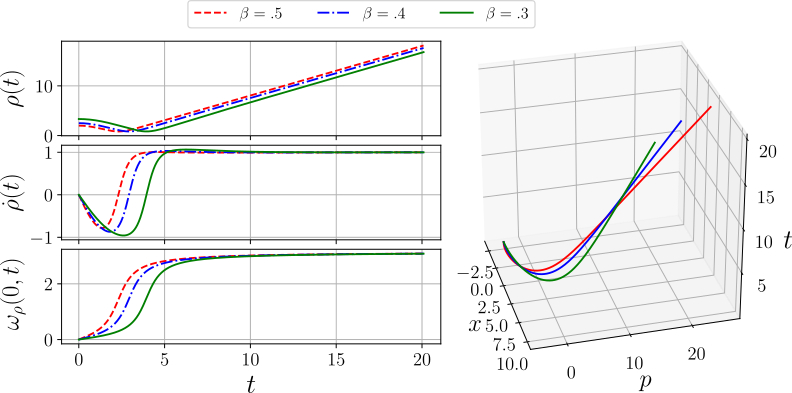}
	\caption{We show the time evolution of the solution to the Ermakov equation $\rho(t)$, its derivative and $\omega_\rho(0,t)$, respectively, for different values of $\beta$. While the 3D graphic shows the time evolution of the mass centre in phase space, where $x_c(0)=-3$ and $p_c(0)=3$ are the position and momentum of the mass centre at time $t = 0$. All variables and constants are in arbitrary units.}
	\label{fig_1}
\end{figure}

Finally, the evolution in phase space of the probability density function $\varrho(x,p;t)$ can be calculated using \eqref{psi_t20} and \eqref{relations_rho}. In figure \eqref{fig_1}, we show the solution of the Ermakov equation, its derivative, $\omega_\rho(0,t)$ and the evolution of the centre of mass in phase space.

\subsection{Oscillatory frequency}\label{s3.2}
\begin{figure}[H]
	\centering
	\includegraphics[width=16cm]{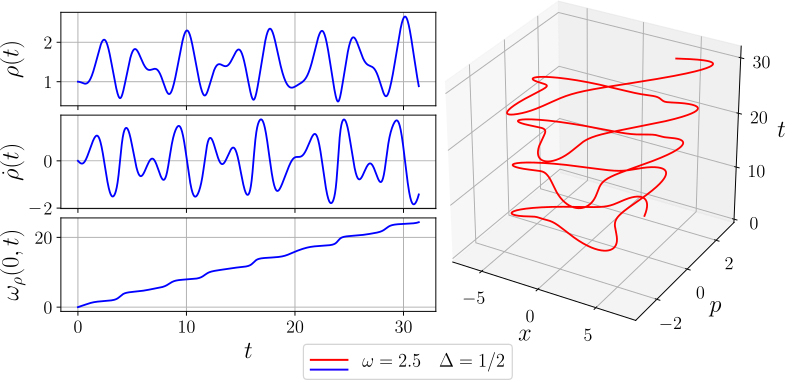}
	\caption{We show the time evolution of the numerical solution of the Ermakov equation $\rho(t)$, its derivative and $\omega_\rho(0,t)$, with $\omega=2.5$ and $\Delta=1/2$. While the 3D graphic shows the time evolution of the mass centre in phase space, where $x_c(0)=2$ and $p_c(0)=2$ are the position and momentum of the mass centre at time $t = 0$. All variables and constants are in arbitrary units.}
	\label{fig_2}
\end{figure}

To conclude this Section we will approach a case that, although it does not have an analytical solution, represents perhaps a more real situation for the time-dependent frequency. The frequency to be considered is
\begin{equation}
k(t)=\Delta+\cos(\omega t).
\end{equation}
To obtain the numerical solution we use the Runge-Kutta method \cite{Johansson_book}. Using the Ermakov equation and defining $x_1(t)=\rho(t)$, $x_2(t)=\dot{\rho}(t)$ and $x_3(t)=\int dt/\rho^2(t)$, we get the system of equations
\begin{align}
\begin{split}
\dot{x}_1(t)&=x_2(t)\\
\dot{x}_2(t)&=\frac{1}{x_1^3(t)}-k(t)x_1(t)\\
\dot{x}_3(t)&=\frac{1}{x_1^2(t)},
\end{split}
\end{align}
where $\rho(0)=1$ and $\dot{\rho}(0)=0$. In figure \eqref{fig_2}, we show the numerical solution of the functions that appear in the state vector \eqref{psi_t20} and the evolution of centre mass in phase space.

\section{Conclusions}
We show that the unification of quantum and classical arguments for dealing with time dependent systems is not only possible, but acquires a simple form with the KvN treatment. The existence of an invariant is always warranted when the Ermakov-Lewis equation is fulfilled for some state vector. Because the operational dynamical modelling is constructed based on the hard-stone theorem of Liouvillan theory in phase space, the representation of the system dynamics turns on a problem easily solved (analytically or numerically) in the position and momentum coordinates over a set of time.

As proved in the examples, when a carefully selected time varying frequency is taken, all the dynamical variables associated with the evolution of the system have a simpler and closed form. From $\rho(t)$ and $\dot{\rho}(t)$ we learned, in the case of subsection Hyperbolically and quadratically growing frequency, that monotonic growing or steady state development can be achieved, while the other case, that of subsection Oscillatory frequency, shows that oscillatory evolution is obtained. In either case, $\omega_{\rho}(0,t)$ tends to grow.

\section{Appendix A}
In general, a Hermitian operator is called invariant if it satisfies the following relationship
\begin{equation}\label{eq_inv}
\frac{\partial\hat{I}}{\partial t}-i[\hat{I},\hat{L}]=0,
\end{equation}
where
\begin{equation}
\hat{L}=\hat{p}\hat{\lambda}_x-U'(\hat{x};t)\hat{\lambda}_p.
\end{equation}
In the case of a time-dependent frequency harmonic oscillator $U'(\hat{x};t)=k(t)\hat{x}$. \\
In order to find the explicit form of the invariant, we write it as
\begin{equation}\label{Invariant}
\hat{I}=\alpha_0(t)\hat{x}^2+\alpha_1(t)\hat{\lambda}_x^2+
\alpha_2(t)\hat{p}^2+\alpha_3(t)\hat{\lambda}_p^2+\alpha_4(t)\hat{x}\hat{p}+\alpha_5(t)\hat{\lambda}_x\hat{\lambda}_p
\end{equation}
and we must find the differential equations that satisfy the coefficients $\alpha_j(t)$. Substituting the invariant \eqref{Invariant} in \eqref{eq_inv} and taking into account the commutation relations
\begin{subequations}\label{alpha_j}
	\begin{align}
	[\hat{x},\hat{p}]=0, \qquad [\hat{\lambda_x},\hat{\lambda_p}]=0, \\
	[\hat{x},\lambda_p]=0, \qquad [\hat{p},\hat{\lambda_x}]=0, \\
	[\hat{x},\lambda_x]=i, \qquad [\hat{p},\hat{\lambda_p}]=i,
	\end{align}
\end{subequations}
we find the system of coupled equations
\begin{subequations}\label{ecs31}
	\begin{align}
	\dot{\alpha}_0-\alpha_4k&=0\label{one},\\
	\dot{\alpha}_1-\alpha_5&=0\label{two},\\
	\dot{\alpha}_2+\alpha_4&=0\label{three},\\
	\dot{\alpha}_3+\alpha_5k&=0\label{four},\\
	\dot{\alpha}_4+2\alpha_0-2 k \alpha_2&=0\label{five},\\
	\dot{\alpha}_5-2\alpha_3+2k \alpha_1&=0\label{six},
	\end{align}
\end{subequations}
where in the sake of simplicity we have remove the time dependence in all variables. 
As result of the symmetry of the commutation relations \eqref{alpha_j}, we get duplicated the same $3 \times 3$ system, one for the even coefficients and one for the odd ones. This system can be reduced to the single equation
\begin{equation}\label{preermakov}
\ddot{\alpha}(t)+2k(t)\alpha(t)=\frac{\dot{\alpha}^2(t)}{2\alpha(t)}+\frac{C}{\alpha(t)}, 
\end{equation}
for $\alpha_2$ or $\alpha_1$, where $C$ is an integration constant. The other even functions are given by
\begin{subequations}\label{coefpares}
	\begin{align}
	\alpha_0&=\frac{\dot{\alpha}_2^2-C_1}{4\alpha_2},\\
	\alpha_4&=-\dot{\alpha}_2,
	\end{align}
\end{subequations}
and the odd ones by
\begin{subequations}\label{coefimpares}
	\begin{align}
	\alpha_3&=\frac{\dot{\alpha}_1^2-C_2}{4\alpha_1},\\
	\alpha_5&=-\frac{\dot{\alpha}_3}{k},
	\end{align}
\end{subequations}
being $C_1$ and $C_2$ integration constants.\\
If in Equation \eqref{preermakov} the solution is proposed as $\alpha_1(t)=\rho^2(t)/2$, we obtain
\begin{equation}
\ddot{\rho}(t)+k(t)\rho(t)=\frac{C}{\rho(t)^3},
\end{equation}
which is the Ermakov equation and  represents the auxiliary condition for the invariant to be found. \\
Taking into account the Ermakov equation and  Equations \eqref{coefpares} and \eqref{coefimpares}, the invariant has the following operational structure
\begin{equation}
\hat{I}=\frac{1}{2}\bigg[\frac{\hat{x}^2}{\rho^2}+(\dot{\rho}\hat{x}-\rho\hat{p})^2+\frac{\hat{\lambda}_p^2}{\rho^2}+(\dot{\rho}\hat{\lambda}_p+\rho\hat{\lambda}_x)^2\bigg].
\end{equation}

\section{Appendix B}
In this Appendix we show how to go from Equation \eqref{psi_t_T} to Equation \eqref{psi_t20}.\\
The first part of Equation \eqref{psi_t_T} contains the term $\hat{T}_2(0)\hat{T}_1(0)\psi(x,p;0)$ and the last part the operator $\hat{T}_1^\dagger(t)\hat{T}_2^\dagger(t)$; thus, we need the following relations
\begin{align}\label{e_kets}
\begin{split}
\exp\left[ \pm if(t)\hat{x}\hat{\lambda}_p\right] 
\hat{p}
\exp\left[ \mp if(t)\hat{x}\hat{\lambda}_p\right] 
&=\hat{p}\pm f(t) \hat{x},\\
\exp\left[ \pm if(t)\hat{p}\hat{\lambda}_x\right] 
\hat{x}
\exp\left[ \mp if(t)\hat{p}\hat{\lambda}_x\right]
&=\hat{x}\pm  f(t) \hat{p},\\
\exp \left[ \pm if(t)(\hat{x}\hat{\lambda}_x+\hat{\lambda}_x\hat{x})\right] 
\hat{x}
\exp \left[ \mp if(t)(\hat{x}\hat{\lambda}_x+\hat{\lambda}_x\hat{x})\right]
&= \exp\left[ \pm2f(t)\right]  \hat{x},\\
\exp \left[ \pm if(t)(\hat{p}\hat{\lambda}_p+\hat{\lambda}_p\hat{p})\right] 
\hat{p}
\exp \left[ \mp if(t)(\hat{p}\hat{\lambda}_p+\hat{\lambda}_p\hat{p})\right]
&=\exp\left[ \pm2f(t)\right]  \hat{p},
\end{split}
\end{align}
with $f(t)$ an arbitrary function. Using these relations in the first part of \eqref{psi_t_T} gives
\begin{align}\label{180223:38}
\hat{T}_2(0)\hat{T}_1(0)\psi(x,p;0)&=\hat{T}_2(0)\psi \left( x,p+\frac{\dot\rho_0}{\rho_0}x;0\right) \nonumber \\
&=\psi \left(\rho_0 x,\frac{p}{\rho_0}+\dot\rho_0 x;0\right) 
\end{align}
where $\rho_0=\rho(0)$ and $\dot\rho_0=\dot\rho(0)$ are the initial conditions of the Ermakov equation.\\
Now, we have to apply the operator  
$ e^{-i\omega_\rho(0,t)( \hat{p}\hat{\lambda}_x-\hat{x}\hat{\lambda}_p)} $
on \eqref{180223:38}; so, we need to disentangle it. In order to do this, we define
\begin{align}
\hat{K}_+=\hat{p}\hat{\lambda}_x,\quad\hat{K}_-=\hat{x}\hat{\lambda}_p,\quad \hat{K}_0=[\hat{K}_+,\hat{K}_-].
\end{align}
As
\begin{equation}
[\hat{K}_\pm,\hat{K}_0]=\pm2\hat{K}_\pm,
\end{equation}
the operators $\hat{K}_\pm$ and  $\hat{K}_0$ are the generators of the angular-momentum algebra and it is very well known \cite{Dattoli} that
\begin{equation}\label{e_des}
e^{-i\omega_\rho(0,t)( \hat{p}\hat{\lambda}_x-\hat{x}\hat{\lambda}_p)}=e^{-i\tan\omega_\rho(0,t)\hat{p}\hat{\lambda}_x}e^{i\ln\cos\omega_\rho(0,t)(\hat{x}\hat{\lambda}_x-\hat{p}\hat{\lambda}_p)}e^{i\tan\omega_\rho(0,t)\hat{x}\hat{\lambda}_p}.
\end{equation}
We have then
\begin{align}\label{psi_t42}
e^{-i\omega_{\rho}(0,t)(\hat{p}\hat{\lambda}_x-\hat{x}\hat{\lambda}_p)}\hat{T}_2(0)\hat{T}_1(0)\psi(x,p;0)=\psi\Big(x\gamma_1-p\gamma_2(t),x\gamma_3(t)+p\gamma_4(t);0\Big)
\end{align}
where
\begin{align}
\begin{split}
\gamma_1(t)&=\rho_0\cos\omega_{\rho}(0,t),\\
\gamma_2(t)&=\rho_0\sin\omega_{\rho}(0,t),\\
\gamma_3(t)&=\frac{\sin\omega_{\rho}(0,t)}{\rho_0}+\dot\rho_0\cos\omega_{\rho}(0,t),\\
\gamma_4(t)&=\frac{\cos\omega_{\rho}(0,t)}{\rho_0}-\dot{\rho}_0\sin\omega_{\rho}(0,t).
\end{split}
\end{align}
Finally, applying again \eqref{e_kets} to the last part of Equation \eqref{psi_t_T}, the state vector at time $t$ is given by
\begin{align}
\psi(x,p;t)=\psi\Big( x\eta_1(t)-p\eta_2(t),x\eta_3(t)+p\eta_4(t);0\Big) 
\end{align}
where
\begin{align}
\begin{split}
\eta_1(t)&=\frac{\rho_0}{\rho(t)}\cos\omega_\rho(0,t)+ \rho_0 \dot{\rho}(t)\sin\omega_\rho(0,t)
,\\
\eta_2(t)&= \rho_0 \rho(t)\sin\omega_\rho(0,t),\\
\eta_3(t)&=\bigg[\frac{1}{\rho_0 \rho(t)}+\dot{\rho}(t)\dot{\rho}_0\bigg]\sin\omega_\rho(0,t)+\bigg[\frac{\dot{\rho}_0}{\rho(t)}-\frac{\dot{\rho}(t)}{\rho_0}\bigg]\cos\omega_\rho(0,t),\\
\eta_4(t)&=\frac{\rho(t)}{\rho_0}\cos\omega_\rho(0,t)-\dot{\rho}_0 \rho(t)\sin\omega_\rho(0,t).
\end{split}
\end{align}

\section*{Acknowledgements}
We are grateful to Denys I. Bondar for his comments and remarks about our manuscript.\\
Irán Ramos-Prieto and Alejandro R. Urzúa-Pineda acknowledge the financial support from Conacyt PhD grants.

\end{document}